%% file: main.tex
\documentclass[reprint,amsmath,amssymb,aps]{revtex4-2}
\usepackage{graphicx} % Include figure files
\usepackage{bm} % bold math
\usepackage[version=4]{mhchem} % chemical formula package
\usepackage{textcomp} % input textcomp
\usepackage{color} % color text
\usepackage{xr}
\usepackage{xcolor,hyperref}
\hypersetup{
   colorlinks,
   linkcolor={blue!50!black},%{red!80!black},
   citecolor={blue!50!black},
   urlcolor={blue!80!black}
}

\usepackage{verbatim}

\newcommand{\modif}[1]{{\color{black}#1}}

\begin{document}

\preprint{APS/123-QED}

\title{\modif{The Decrease of Static Friction Coefficient with Interface Growth from Single to Multi-asperity Contact}}% Force line breaks with \\

\author{Liang Peng $^{1}$$^{*}$}
\author{Thibault Roch$^{1}$$^{*}$} 
\author{Daniel Bonn$^{1}$} 
\author{Bart Weber$^{1,2}$} 
\affiliation{$^{1}$ Van der Waals-Zeeman Institute, Institute of Physics, University of Amsterdam, Science Park 904, 1098 XH Amsterdam, The Netherlands \\
$^{2}$ Advanced Research Center for Nanolithography (ARCNL), Science Park 106, 1098 XG Amsterdam, The Netherlands}

%TC:ignore
\begin{abstract}
The key parameter for describing frictional strength at the onset of sliding is the static friction coefficient. Yet, how the static friction coefficient emerges at the macroscale from contacting asperities at the microscale is still an open problem. Here, we present friction experiments in which the normal load was varied over more than three orders of magnitude, so that a transition from a single asperity contact at low loads to multi-asperity contacts at high loads was achieved. We find a remarkable reduction in the friction drop (the ratio of the static friction force to the dynamic friction force) with increasing normal load. Using a simple stick-slip transition model we identify the presence of \mbox{pre-sliding} and subcritical contact points as the cause of smaller static friction coefficient at increased normal loads. Our measurements and model bridge the gap between friction behavior commonly observed in atomic force microscopy (AFM) experiments at microscopic forces, and industrially relevant \mbox{multi-asperity} contact interfaces loaded with macroscopic forces.
\end{abstract}
%TC:endignore

% Don't count these!
%TC:ignore
%\quickwordcount{Main}
%\quickcharcount{Main}
%\detailtexcount{Main}

\maketitle
%TC:endignore

The onset of sliding is of critical importance to various natural and industrial interfaces. For example, stick-slip friction behavior leads to catastrophic energy release in earthquakes \cite{ruina1983, Scholz1998, ikari2013, farain2024} and can hamper the functioning of microelectromechanical systems (MEMS) \cite{corwin2004, weber2022,shroff2014rate}. When slip is initiated, the static friction force generically reduces to a (lower) dynamic friction force \cite{weber2019, baumberger2006solid, mate2019}. This transition from static to dynamic friction, also known as friction drop, can be effectively described by the empirical rate and state friction (RSF) laws across various contact length scales \cite{li2020, tian2019, frerot2021, sahli2018}. However, the underlying physical mechanisms remain elusive.

Atomic force microscopy (AFM) friction experiments \cite{li2011, tian2017, vorholzer2019, badt2024} and theoretical models \cite{ouyang2019, liu2012, filippov2004, li2022} have shown that at silicon/silicon contacts, the dynamics of the formation and rupture of interfacial bond networks control the evolution of friction during the onset of slip. While recent experiments suggest that similar bonding phenomena also occur at larger multi-contact interfaces \cite{Peng2023}, additional complexity may emerge as contacts grow \cite{li2020}. Examples of such complexity include the wear \cite{leriche2023atomic}, \mbox{multi-scale} roughness \cite{goebel2023fault}, the widening of interfacial pressure distribution \cite{Hsu2022}, nonconcurrent slip across the interface \cite{rubinstein2004detachment,ben2010slip,kammer2012propagation} and interactions between neighboring contact asperities \cite{Li2018Chemical, leriche2023atomic}. Systematic studies into how such complexity emerges as contacts grow are challenging because rough macroscopic contact interfaces consist of a myriad of contact asperities that are hidden from view by the bulk contacting bodies \cite{ben2010slip, weber2018, romero2014}.

In this Letter, we experimentally demonstrate that the \modif{friction drop (the ratio of static to dynamic friction coefficient)} decreases as the interface grows from single contact to multi-asperity contact by increasing the normal load applied to a silicon-on-silicon contact. The observed decreasing trend of \modif{friction drop} with increasing normal load can be captured by a simple stick-slip transition model, analogous to a fiber bundle model, based on the local normal contact stress map obtained from boundary integral contact calculations without adjustable parameters. We find that three types of contact points can be identified when the interface is subjected to the maximal tangential force it can withstand. We describe these three types of contact elements as ``critical'', ``\mbox{pre-sliding}'' and ``subcritical''. While single-asperity interfaces are always critical at the onset of slip, we show that the emergence of \mbox{pre-sliding} and subcritical contact asperities at multi-asperity interfaces causes the \modif{friction drop} to decrease.

\begin{figure}[ht]
    \centering
    \includegraphics[width=0.5\textwidth]{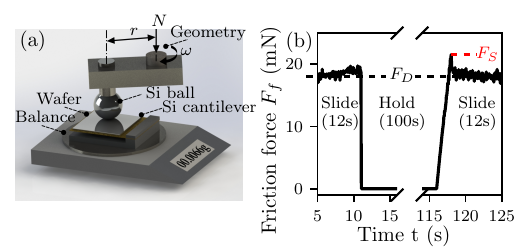}
    \caption{Experimental system. (a) A 3 \mbox{mm-diameter} silicon ball is clamped to the geometry of a rheometer and brought into contact with a homemade Si cantilever. By rotating the geometry at a constant angular velocity, $\omega$, a sliding speed of $\omega r$  ($r=10 \text{ mm}$) is imposed at the Si-on-Si contact interface. The corresponding friction force ($F_f$) and the normal force ($N$) can be simultaneously measured by the rheometer and the balance. The ratio of friction force and normal load gives the coefficient of friction, $\mu ={F_f}/{N}$. (b) {An example of the slide-hold-slide experiments protocol with a normal load of 24 mN and a hold time of 100 s.} $F_S$ and $F_D$ represent the maximum static friction force and the average dynamic friction force after the sliding is reinitiated, respectively.}
    \label{fig:fig1}
\end{figure}

We study silicon-on-silicon friction using a rheometer (DSR 502, Anton Paar) in combination with an analytical balance (Quintix, Sartorius) inside a \mbox{humidity-controlled} chamber, see Fig.~\ref{fig:fig1}(a). To prepare highly hydroxylated silicon surfaces \cite{tian2017, li2011}, the silicon balls and the silicon cantilevers were treated by oxygen plasma cleaning inside a reactor (Zepto One, Diener Electronic) \cite{Peng2023}. In the friction experiments, the normal loads were precisely varied from 0.1 to 100 \text{mN} by using silicon cantilevers of different normal stiffness (see more details in Appendix~\ref{sec:supp_exp} \nocite{socoliuc2006atomic}). The imposed sliding speed was fixed at 0.3 \,\textmu m/s for all the measurements. To reset contact aging \cite{rubinstein2006, vorholzer2019,tian2017}, a \mbox{slide-hold-slide} (SHS) measurement procedure was employed (see Fig.~\ref{fig:fig1}(b)): we used the rheometer to impose a stationary state with zero-torque (or zero-friction) after the initial sliding to avoid impacting the aging process with shear force \cite{dillavou2020shear,farain2024, vorholzer2019}, followed by reinitiating the sliding in the same direction. The static friction force $F_S$ and dynamic friction force $F_D$ when sliding was restarted, were recorded to calculate the corresponding static friction coefficient, $\mu_S$, the dynamic friction coefficient, $\mu_D$, and the friction drop, $\mu_S/\mu_D$, which equals to $F_S/F_D$. All sliding strokes were measured on previously untouched areas of the silicon wafer and kept short at 10 \textmu m to minimize the influence of wear (Fig.~\ref{fig:figs2}).
 
\begin{figure}[h]
    \centering
    \includegraphics[width=0.5\textwidth]{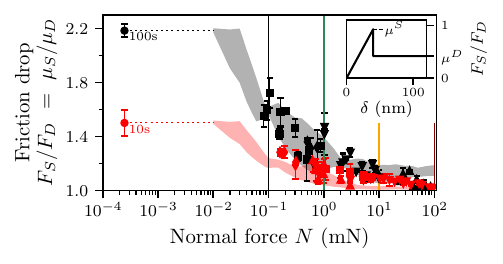}
    \caption{The friction drop as a function of normal load, measured from experiments (data points) and predicted by our model (shaded area), with black and red corresponding to a hold time of 100 s and 10 s, respectively \modif{(the static and dynamic friction coefficients are reported in Fig.~\ref{fig:figs1}.)}. Each symbol stands for a different Si cantilever with varying normal stiffness, and each data point represents the average of three measurements. The leftmost red and black solid circle data points, measured from AFM experiments, are extracted from Fig.~S5 in Ref. \cite{li2011}, with the evolution of the nanoscale friction coefficient being reproduced in the inset. The vertical lines indicate the normal load corresponding to the normal stress distribution shown in Fig.~\ref{fig:fig3}.}
    \label{fig:fig2}
\end{figure}

Both the static and the dynamic friction coefficients, defined as the ratio of friction force to normal force, are widely assumed to be independent of the normal load \cite{berthoud1999physical,baumberger2006solid,capozza2012static}. Surprisingly, we observe that the \modif{friction drop} gradually decreases with increasing normal load as shown with black data points in Fig.~\ref{fig:fig2}. Notably, the nanoscale single-asperity friction drop (leftmost black circle in Fig.~\ref{fig:fig2}), reproduced from previous AFM experiments (inset of Fig.~\ref{fig:fig2}) \cite{li2011}, aligns well with our observed qualitative decreasing trend of friction drop with increasing normal load.

To provide insight into the normal load dependence of the friction drop, we propose a simple stick-slip transition model. \modif{In the model, we discretize the interface with square elements (pixels) such that the macroscopic friction force is the sum of all the microscopic interfacial shear forces generated by these elements.} We also assume that the \modif{element-scale} frictional behavior is characterized by an instant drop in friction upon slip, the magnitude of which matches that observed in the AFM friction experiments shown in the inset of Fig.~\ref{fig:fig2}. In other words, the local ratio of shear stress to normal stress \modif{(local stress ratio)} drops at the onset of local slip, in accordance with the static and dynamic friction coefficient ($\mu^S$ and $\mu^D$, respectively) observed in the AFM experiments. \modif{The interface is a collection of microscopic contact patches of varying size (see Fig.~\ref{fig:figs4}). In our model, the shear force applied to an individual contact patch scales linearly with the number of elements involved in that patch. This means that the applied shear force is distributed homogeneously \modif{over the real contact area} before any element starts to slip}. The excess shear force generated when an element enters slip is uniformly redistributed over \modif{the interfacial contact area (elements)} that has not undergone slip yet. \modif{We also assume that there is no evolution of the area and geometry of the contact patches during sliding, as the substrate wafer is much smoother than the silicon ball \cite{Peng2023}.}

\begin{figure}[ht]
    \centering
    \includegraphics[width=0.45\textwidth]{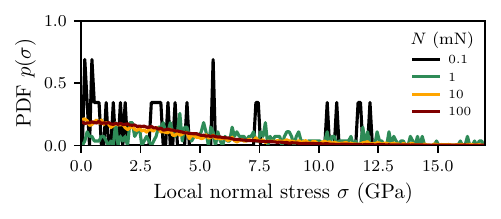}
    \caption{The probability density function, \( p(\sigma) \), of the local normal stress \( \sigma \) under various normal loads.}
    \label{fig:fig3}
\end{figure}

To estimate the local normal stress at the interface, we conduct elastic contact calculations using an open-source implementation of the boundary integral method (BIM) \cite{Frerot2020}. \modif{In the contact calculation, the elastically stored energy of the contact system is minimized under the constraint that the resulting mean contact pressure balances the applied average pressure over the area of the input topography (see more details in Appendix~\ref{sec:supp_numerical}} Our use of elastic contact is reasonable here, since there is mainly elastic deformation at the contact interface (see Fig.~\ref{fig:fig3}) \cite{Li2018Chemical,Peng2023,hsia2022contribution}. Using the calculated local normal stress distribution and the assumed relations between local shear stress and local normal stress described above, the model predicts how the total tangential force, which is externally imposed, is distributed over the contact area and how this distribution of local shear stress evolves as the applied tangential force is increased. \modif{In our model, at the level of a single element, the maximal (critical) ratio of shear stress to normal stress is constant; in line with the load-controlled friction framework \cite{berman1998amontons,gao2004frictional,Hsia2021}.} It is the normal stress distribution that completely determines the macroscopic frictional behavior. This enables us to predict the frictional response of the same interface at various normal loads, from which we can compute the macroscopic static friction coefficient and friction drop.

In other words, our model is analogous to a fiber bundle model with equal load sharing \cite{pradhan_failure_2010}, where the failure of a bundle of fibers corresponds to the failure of contact elements, with their strength governed by the local normal stress. While sliding heterogeneity can originate from elastic interaction among contact patches \cite{li2020}, we assume that these interactions during the onset of the sliding process are negligible in our system. Indeed, the nanometer-scale contact deformation ($\sim$ 5 nm at a normal load of 10 mN based on BIM contact calculation as mentioned above) is considerably smaller than the \mbox{micrometer-scale} distance between contact patches. \modif{Therefore, it is reasonable to ignore the role of asperity interactions in the distribution of the shear forces. However, this assumption may break down at increasing normal loads when the distances between contact patches shrink (see Fig. S4).}

To quantify ball-to-ball topography variations, we use three different topographies measured on silicon balls that were used in the experiments (see Fig.~\ref{fig:figs3}). For each topography and normal load, the model predicts the friction drop. The gray shaded area in Fig.~\ref{fig:fig2} corresponds to the range of predicted friction drop values based on the three topographies, and the nanoscale static and dynamic friction coefficients measured at a hold time of 100 s in AFM experiments. The stick-slip transition model qualitatively reproduces the decrease of the friction drop with increasing normal load and quantitatively matches the experiments without any fitting parameter.

To evaluate the robustness of our model against variations in the system characteristics, we conduct additional experiments with a hold time of 10 s instead of 100 s, and report the corresponding friction drop with red data points in Fig.~\ref{fig:fig2}. A decrease of the friction drop with increasing normal load is observed, similar to the experiments for a hold time of 100 s. Furthermore, it is noteworthy that the measured friction drop for a hold time of 100 s is larger than that for a hold time of 10 s under the same normal force. The larger static friction coefficient, corresponding to a greater friction drop, is expected: a longer hold time during the SHS procedure can enhance the contact strength by forming more interfacial siloxane bonds \cite{li2011,Peng2023}, which results in a higher static friction coefficient. Moreover, our stick-slip transition model (red-shaded area in Fig.~\ref{fig:fig2}) with the nanoscale static and dynamic friction coefficients measured at a hold time of 10 s from previous AFM experiments \cite{li2011} (leftmost red circle in Fig.~\ref{fig:fig2}, see also Fig.~\ref{fig:figs5}), still captures successfully the evolution of friction drop with the normal force for the experiments with 10 s hold time. One should note that there is still a gap between the minimum normal load used in the simulation and AFM experiments in Fig.~\ref{fig:fig2} (marked by black and red horizontal dot lines), which will be discussed later.

To further elucidate the relation between the normal load, the local normal stress distribution, and the macroscopic frictional response, we plot the probability density function,  \( p(\sigma) \), of the normal stress at the interface, obtained from elastic contact calculations with various normal loads in Fig.~\ref{fig:fig3}. With increasing normal load, the probability density function of the local normal stress evolves from being discrete (the black curve, 0.1 mN) to continuous (the yellow curve, 10mN). This directly relates to the transition from single to multi-asperity contact interfaces \modif{(see Fig.~\ref{fig:figs4})}. At lowest normal loads, only a few contact patches are formed, and the macroscopic behavior may be dominated by those patches. At larger normal loads, however, there is a more continuous distribution of normal (and thus shear) stresses at the interface. The normal loads corresponding to the distribution shown in Fig.~\ref{fig:fig3} are indicated with vertical lines of the same color in Fig.~\ref{fig:fig2}. 

\begin{figure*}[ht]
    \centering
    \includegraphics[width=0.9\textwidth]{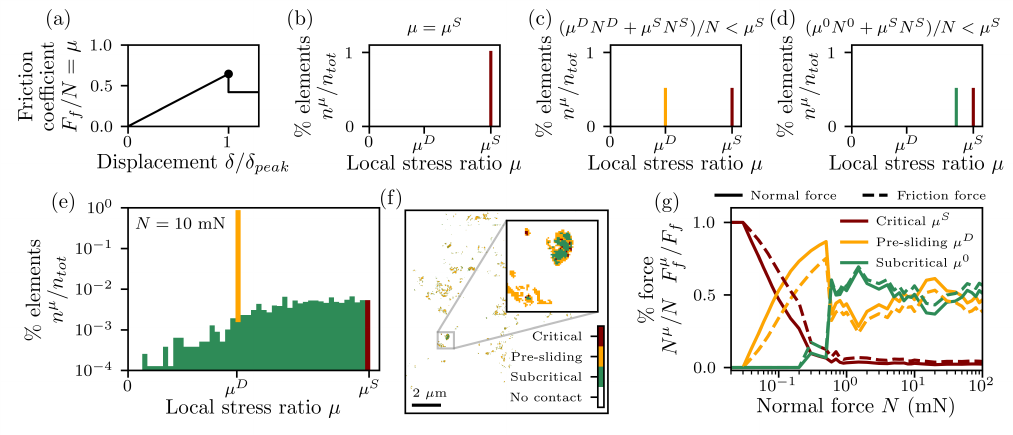}
    \caption{Local stick-slip transition at the critical state of the interface from numerical computation. (a) Macroscopic friction coefficient versus displacement curve at a normal load of 0.1 \text{mN}. The black circle corresponds to the critical interfacial state where no more additional tangential load can be added. (b) to (e) Distribution of \modif{local stress ratio} for a single-element interface (b), \mbox{two-element} interface (c)(d), and \mbox{multi-element} interface under a normal load of 10 \text{mN} (e), respectively. The brown, yellow, and green correspond to critical, pre-sliding, and subcritical elements with \modif{local stress ratios} of $\mu^D$, $\mu^S$, and $\mu^0$, respectively. $\mu^0$ is smaller than $\mu^S$. $n^\mu$ and $n^{tot}$ represent the number of elements under a \modif{local stress ratio} of $\mu$ ($\mu = \mu^S$, $\mu^D$, or $\mu^0$) and the total number of elements, respectively. Movies of the evolution of the distribution of \modif{local stress ratios} with increasing imposed displacement are available in Appendix~\ref{sec:supp_movies}. (f) The state of the contacting elements at the critical interfacial state for a normal load of 10 \text{mN}. (g) The fraction of the total normal force (solid lines) and the total friction force (dashed lines) generated by different types of contact elements in the interfaces under various normal loads. $N^\mu$ and $N$ represent the shared normal load on those contact elements under a \modif{local stress ratio} of $\mu$ ($\mu = \mu^S$, $\mu^D$, or $\mu^0$) and the total applied normal load, respectively. $F_f^\mu$ and $F_f$ represent the same quantity but for the shear load.} 
   \label{fig:fig4}
\end{figure*}   

To better understand how the differences in normal stress distribution impact the onset of sliding, we focus on the critical state of the interface, where the peak macroscopic friction force is reached under an imposed tangential lateral displacement of $\delta_{\text{peak}}$ (see a typical simulated friction-displacement curve in Fig.~\ref{fig:fig4}(a)). We plot the distribution of \modif{local stress ratios} across all the contact elements at the critical state of the interface in Fig.~\ref{fig:fig4}(b)-(e). For an \modif{idealized} single element contact (Fig.~\ref{fig:fig4}(b)), the element will reach the maximum \modif{local stress ratio}, $\mu^S$, resulting in an interfacial friction coefficient equal to the \modif{local stress ratio} ($\mu=\mu^S$). In this limiting case of a single asperity contact, one recovers the frictional behavior as measured from AFM. In our model, this means that below a normal load for which the real contact area is of the scale of the input AFM topography resolution (21 \text{nm}), the macroscopic friction coefficient is not dependent on the normal load. Such an extreme case is marked by black and red horizontal dot lines in Fig.~\ref{fig:fig2}.

The situation becomes more complex when considering \mbox{multi-asperity} contact interfaces, as each element potentially experiences different local normal stresses (and thus different \modif{local stress ratios} at the critical interfacial state). We stress that, within our model, any distribution of \modif{local stress ratios} (between $0$ and $\mu^S$) at a multi-contact interface can only result in a decrease in the friction drop as compared to the friction drop at single-asperity level. Hereafter we investigate the details of the critical interfacial state for multi-asperity interfaces and its influence on the friction drop, \modif{starting with the idealized case of two contacting elements.} 

For an interface composed of two contacting elements, each carrying a different normal stress, two distinct scenarios arise at the critical interfacial state corresponding to maximal (global) static friction force. In the first scenario, the normal stress carried by each element varies significantly (Fig.~\ref{fig:fig4}(c)) such that the weaker element, carrying a smaller normal stress, has already slipped and experiences \modif{local stress ratio} $\mu^D$, while the stronger element with the higher normal stress remains stuck with a \modif{local stress ratio} of $\mu^S$. We can calculate the interfacial friction coefficient as $\mu=(\mu^DN^D+\mu^SN^S)/N$, with $N^D$ and $N^S$ the normal forces carried by the two elements: $N=N^D+N^S$. In the second scenario, the difference in the normal stresses carried by the two elements is smaller (Fig.~\ref{fig:fig4}(d)). This results in both contact elements being stuck at the critical state of the interface. Now, the failure of the weaker element will trigger the slip of the stronger element due to its excess friction force redistribution. \modif{This is reminiscent of the asperity model in geophysics \cite{johnson2002asperity}, in which the failure of one asperity may cause an avalanche of asperity failure}. The element with lower normal stress has reached a \modif{local stress ratio} of $\mu^S$, while the contact element with higher normal stress experiences a local stress ratio of $\mu^0$ that is smaller than $\mu^S$. Contrarily to $\mu^D$ and $\mu^S$, $\mu^0$ is not a fixed value related to the nanoscale friction experiment: it is the instantaneous \modif{local stress ratio} experienced by a stuck element that is loaded sub-critically. The collective friction coefficient is then $\mu=(\mu^0N^0+\mu^SN^S)/N$, where $N^0$ and $N^S$ represent the normal force on the two elements, satisfying $N=N^0+N^S$ and $N^0>N^S$. In both scenarios the friction drop is lowered by the addition of a second element, either because an element has already entered slip when the overall friction force is maximal (scenario 1) or because the stronger element is ``subcritical''; its slip is activated by failure of the weaker element (scenario 2). 

We can therefore define three types of contact elements at the critical state of \mbox{multi-asperity} interfaces: critical contact elements, experiencing the maximal \modif{local stress ratio} $\mu^S$; pre-sliding contact elements, that have already slipped and therefore show the dynamic \modif{stress ratio} $\mu^D$;  and subcritical contact elements, with a local stress ratio of $\mu^0$ smaller than $\mu^S$. For a realistic \mbox{multi-asperity} interface, many elements might correspond to each of the three element types. An example is shown in Fig.~\ref{fig:fig4}(e), with the distribution of local stress ratios for an interface at the critical state under a normal load of 10 \text{mN}. The three types of elements (critical, pre-sliding, and subcritical) are respectively represented in green, yellow, and brown. The corresponding contact map from the BIM computation is shown in Fig.~\ref{fig:fig4}(f), with the contact being colored following the same scheme.

The slip of the subcritical elements is activated by the redistribution of excess friction force released when critical elements slip. This instability in our model defines the critical interface state with maximal applied tangential load. Since both $\mu^D$ and $\mu^0$ are smaller than $\mu^S$, the collective friction coefficient at the critical interfacial state will, as expected, be smaller as compared to that in a single element interface ($\mu<\mu^S$). Consequently, it is the pre-sliding and subcritical contact elements that dominate the reduction of the collective static friction coefficient at the \mbox{multi-asperity} interfaces with respect to its value for a single element interface, leading to a smaller friction drop with increasing normal load. To quantify the relative contribution of critical, pre-sliding, and subcritical contact elements to the overall static friction coefficient, we plot the fraction of the total normal load (respectively friction force) supported by different types of contact elements under various total normal loads in solid (respectively dashed) lines in Fig.~\ref{fig:fig4}(g). As the total normal load increases, more contact elements form at the contact interface. The fraction of pre-sliding and subcritical contact elements gradually increases from 0 at single element contact interfaces to an approximately constant value at \mbox{multi-asperity} contact interfaces, which suggests that the friction drop decreases with increasing normal force. In contrast, the fraction of critical contact elements decreases from 1 to an almost negligible level. This decreasing fraction of critical contact elements with increasing normal load indicates a smaller contribution of critical contact elements to the static friction coefficient, which matches well with the observed friction drop in Fig.~\ref{fig:fig2}. \modif{In the low normal load limit, the evolution of friction drop is given by the intrinsic, nano-scale friction drop which can for example be controlled by interfacial bonding dynamics \cite{leriche2023atomic}. At low normal loads, the critical asperities govern the static friction coefficient, which is analogous to the dominant role of nucleation regions as identified in previous work \cite{dillavou2018,rubinstein2007}. The specific normal load range over which the friction drop decreases, is controlled by the elastic properties and surface roughness in our model, with the relation between friction drop and normal load shifting to higher normal loads for higher roughness or stiffer materials as long as the contact mechanics remains elastic. For example, MEMS devices typically involve lower roughness and lower normal force compared to our experimental conditions, and may therefore display similar friction drop behavior. At larger interfaces, however, additional complexity may arise \cite{rubinstein2004detachment,goebel2017allows}.} \modif{The assumption of independent contact elements in our model may break down at larger loads, where large contact patches lead to increased interfacial stiffness and shear loading \cite{mindlin1949compliance}, resulting in an even more heterogeneous distribution of shear stress, which could further reduce the predicted friction drop.}

In conclusion, we have investigated how the macroscopic frictional response of a \mbox{Si-on-Si} system evolves from a single to a multi-asperity contact interface, and show how the macroscopic static friction coefficient decreases with increasing normal load across a broad range of loads. We propose a stick-slip model, analogous to a fiber bundle model, based on elastic contact calculations to capture the decrease in macroscopic static friction coefficient and friction drop with increasing normal load. This decrease is found to be dominated by the relative fraction of critical contact elements at the interface that collectively fail when the maximal static friction force is reached. With increasing normal load, we observe the growing populations of pre-sliding elements -contact elements that slip before the maximal tangential force is applied to the interface- and subcritical elements -contact elements whose slip is activated by redistribution of the excess friction from critical elements. Both the pre-sliding and subcritical populations of elements observed at the critical interface state grow with increasing normal load, causing the friction drop to decrease with increasing normal force. By changing the hold time in our hold-slide-hold experiments, we demonstrated that these results remain valid when considering a different reference nanoscale friction behavior. All in all, our findings shed new light on the complexity that emerges when contact interfaces scale up from single asperity to multi-asperity - a central question in tribology and an important challenge in precision positioning.

%TC:ignore
\begin{acknowledgments}
We thank R. Carpick, B. Mcclimon, J. Dijksman and K. Farain for valuable discussions. TR acknowledges support of the Swiss National Science Foundation through the fellowship No. P500PT/222342. LP acknowledges funding from the China Scholarship Council and thanks the invaluable support from Min Li.
\end{acknowledgments}

\thanks[* L. Peng and T. Roch contributed equally.
\bibliography{Reference}

\input{Supplemental_material}

\end{document}

%% file: Supplemental_material.tex
\clearpage
\newpage
\pagebreak
\appendix*
\setcounter{figure}{0}
\renewcommand{\thefigure}{S\arabic{figure}}

{\linespread{1.35}\section{{Supplemental Material for: The Decrease of Static Friction Coefficient with Interface Growth from Single to Multi-asperity Contact}}}
\renewcommand{\thesubsection}{\Alph{subsection}}

\subsection{Experimental Method}
\label{sec:supp_exp}

The rheometer (DSR 502, Anton Paar) and an analytical balance (Quintix, Satorious) were used to conduct the friction measurements. 
A 3 \mbox{mm-diameter} silicon ball was clamped to the geometry of a rheometer and brought into contact with a homemade Si cantilever that was made by gluing a piece of 1×1 cm p-type silicon (100) wafer (University Wafer) to a brass sheet. The normal stiffness of the Si cantilever could be tuned by changing the width and thickness of the used brass sheet. The Si cantilever was mounted on a stage, which was attached to the weighing pan of an analytical balance by dead weight. Before the friction experiments, the (naturally oxidized) silicon balls (Goodfellow) were sonicated in ethanol and then \mbox{Milli-Q} water, followed by nitrogen flow drying. To compare with the highly hydroxylated silicon surfaces used in previous AFM experiments \cite{tian2017, li2011}, the silicon balls and the silicon cantilevers were treated by oxygen plasma cleaning inside a reactor (Zepto One, Diener Electronic) \cite{Peng2023}. In the friction experiments, both the plasma-cleaned silicon ball and the plasma-cleaned silicon cantilever were initially equilibrated for one hour inside the dry chamber before contact was formed between them. The relative humidity inside the chamber was stabilized at 0.8\% by continuously flowing dry nitrogen. During sliding, the rheometer and balance simultaneously measured the friction force ($F_f$) and normal load ($N$) at a sampling rate of 20 Hz, corresponding to one data point for every 15 nm of displacement. When the sliding was restarted, we reported the peak static friction ($F_S$ or $\mu_S$) and the average value of the dynamic friction ($F_D$ or $\mu_D$) in the steady state sliding regime, measured over a displacement of  \(\sim 1 \, \mu\text{m} \) after the peak friction. We show in Fig.~S1 the static and dynamic friction coefficients at a hold time of 100 s.

\begin{figure}[th!]
    \includegraphics[width=0.45\textwidth]{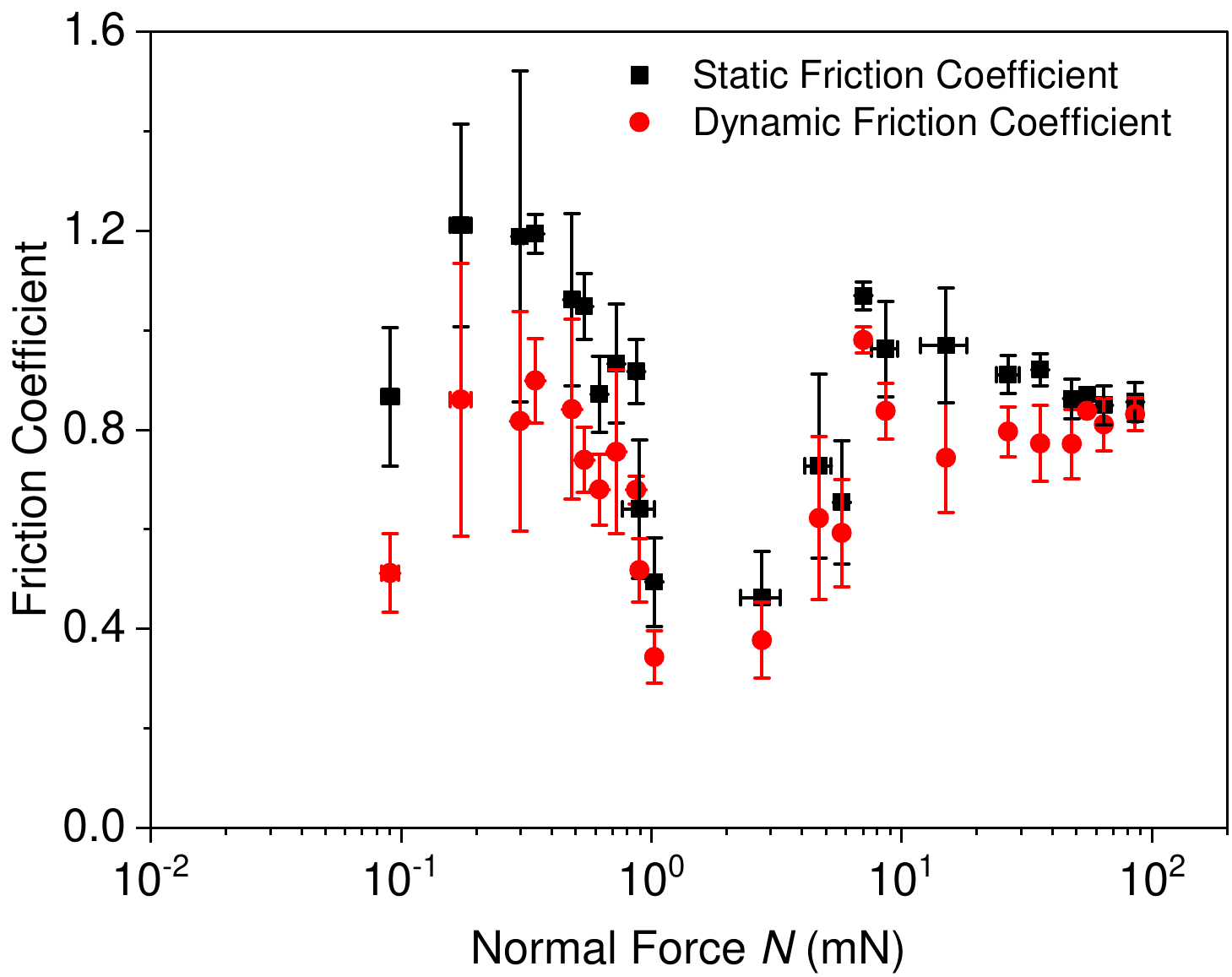}
    \caption{Friction coefficient as a function of normal load at a hold time of 100 s. The black and red data correspond to the static and dynamic friction coefficient, respectively. The measured static friction coefficient shows a decreasing trend with increasing normal load, while the dynamic friction coefficient remains relatively constant under the same conditions. Notably, a dip in both static and dynamic friction coefficients is observed around a normal load of 1 mN, which does not affect the friction drop reported in the main text, as both coefficients contribute equally to the ratio. We suspect that the dip is caused by a resonance in the system, as oscillating normal force is known to decrease friction \cite{socoliuc2006atomic,farain2024perturbation}. The contact stiffness decreases with decreasing normal force and may enable vibrations in the normal direction, thereby lowering both static and dynamic friction, at around 1 mN.}
    \label{fig:figs1}
\end{figure}

The application of shear force from above (and below) the interface can cause torque in friction experiments, which may impact the distribution of normal stress across the interface. In flat-on-flat geometries, this effect has been utilized to ensure that the onset of failure is located at the leading or trailing edge of the interface \cite{dillavou2020shear}. In the sphere-on-flat experimental geometry we use here, the redistribution of the interfacial normal stress due to torque is negligible \cite{weber2019}.

\begin{figure*}[th!]
    \includegraphics[width=0.9\textwidth]{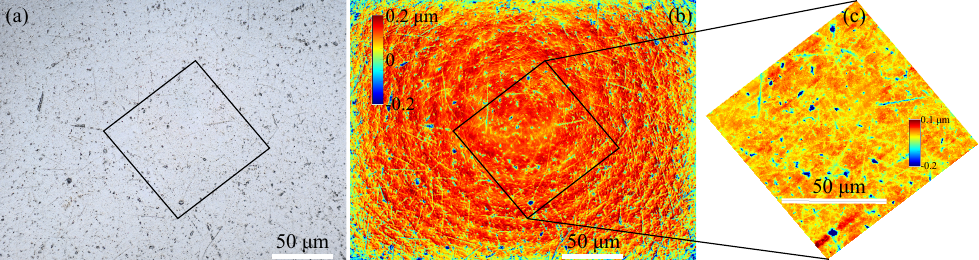}
    \caption{Topography of silicon ball apex after friction experiments. (a) Optical image of the apex of silicon ball captured over an area of 283 $\times$ 212 $\mu$m\textsuperscript{2}, using a laser-scanning confocal profilometer (Keyence VK-X1000). (b) The height profile derived from the optical image in (a). (c) AFM topography with a size of 90 $\times$ 90 $\mu$m\textsuperscript{2}, which corresponds to the square-marked area in (a) and (c). Both optical and AFM images reveal no visible signs of wear, indicating the minimal impact of wear on our experimental result. Scale bar, 50 $\mu$m.}
    \label{fig:figs2}
\end{figure*}

\begin{figure*}[th]
    \includegraphics[width=0.9\textwidth]{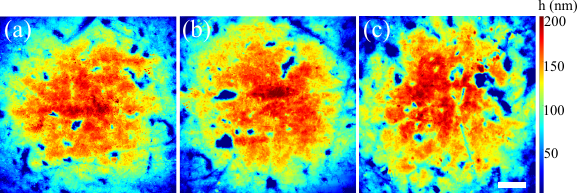}
    \caption{AFM topography. (a)(b)(c) Three different AFM topographies measured from silicon balls used in experiments. Measurements were taken using tapping mode in AFM (Dimension Icon, Bruker) with Si tips (RTESPA-300, Bruker). The topography size is 31.13 × 31.13 $\mu\text{m}^2$ with an RMS (root-mean-square) roughness around 40 nm. Scale bar, 5 $\mu$m..}
    \label{fig:figs3}
\end{figure*}

\subsection{Numerical Method}
\label{sec:supp_numerical}

We use an open-source implementation of the boundary integral method \cite{Frerot2020} for the elastic contact calculation. The contact between a silicon ball and a silicon wafer is modeled as a single rough surface in contact with a rigid flat surface, with the roughness from both silicon surfaces mapped onto the single rough surface. The contact problem was then solved with a compound roughness $h = h_1 - h_2$, with $h_1$ the roughness of the silicon ball and $h_2$ the roughness of the silicon wafer measured with AFM after friction experiments, see Fig.~S2 and Fig.~S3. The surface topography size is 31.13 × 31.13 $\mu\text{m}^2$ and is made of 1500 × 1500 elements, each with a length of 20.75 × 20.75 nm. We use the mechanical properties of silicon: Young's modulus: \(E = 130 \pm 3 \, \text{GPa}\), Poisson's ratio: \(\nu = 0.26\). The apparent pressure is imposed and varied to get the contact pressure map at various total normal loads. Examples of the contact maps under various loads are presented in Fig.~S4, illustrating the evolution from single to multi-contact asperity interface. For $N=0.1$mN, the contact interface is made of three continuous contact patches (each made of several elements), see the inset in Fig.~S4(a). The total contact area in this case spans around 50 elements. Two of these contact patches eventually merge to form a single contact when the normal load is increased to $N=1$mN, accompanied by the formation of a new contact patch at another location of the interface, see Fig.~S4(b). The map of the contact under a load of $N=10$mN (Fig.~S4(c)) is radically different, with the contact made of numerous contact patches (each of them several elements in size).

\begin{figure}[th!]
    \centering
    \includegraphics[width=0.45\textwidth]{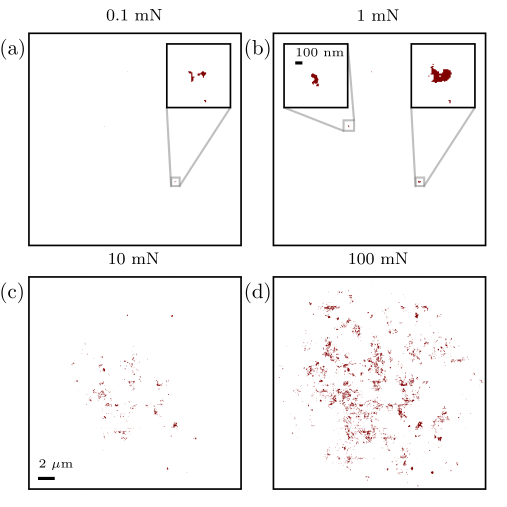}
    \caption{{Contact maps under normal loads of (a) $N=0.1$mN, (b) $N=1$mN, (c) $N=10$mN and (d) $N=100$mN. The inset in panel (a) zooms into the only area exhibiting contact under this load. The insets in panel (b) show the two main areas of contact.}}
    \label{fig:figs4}
\end{figure}

\begin{figure}[th!]
    \centering
    \includegraphics[width=0.45\textwidth]{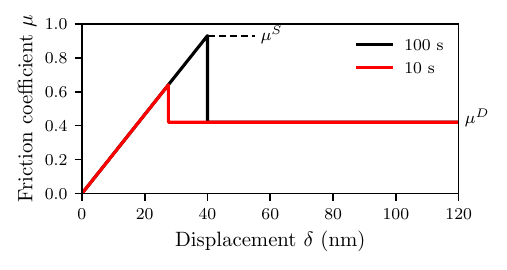}
    \caption{Nanoscale friction coefficient-displacement curve used for simulation in Fig.~2. The black and red curves correspond to the AFM SHS measurements for a hold time of 100 s and 10 s, respectively. The static and dynamic friction values for a hold time of 100 s are indicated respectively with $\mu^S$ and $\mu^D$. The nanoscale friction coefficient is calculated as the ratio of the friction force to the total normal force, comprising the adhesion force and the externally applied load.}
    \label{fig:figs5}
\end{figure}

For the stick-slip model, the friction behavior at the element level follows the measurements from AFM experiments \cite{li2011}. The behavior for 100 s and 10 s waiting times are shown in Fig.~\ref{fig:figs5}.

\subsection{Supplemental movies}
\label{sec:supp_movies}

Three supplemental movies are associated with this manuscript:

\begin{itemize}
    \item ``SupplementalMovie01mN.mp4'', focusing on a normal load $N = $ 0.1 mN.
    \item ``SupplementalMovie1mN.mp4'', focusing on a normal load $N = $ 1 mN.
    \item ``SupplementalMovie10mN.mp4'', focusing on a normal load $N = $ 10 mN.
\end{itemize}

\begin{figure*}[th!]
    \centering
    \includegraphics[scale=1]{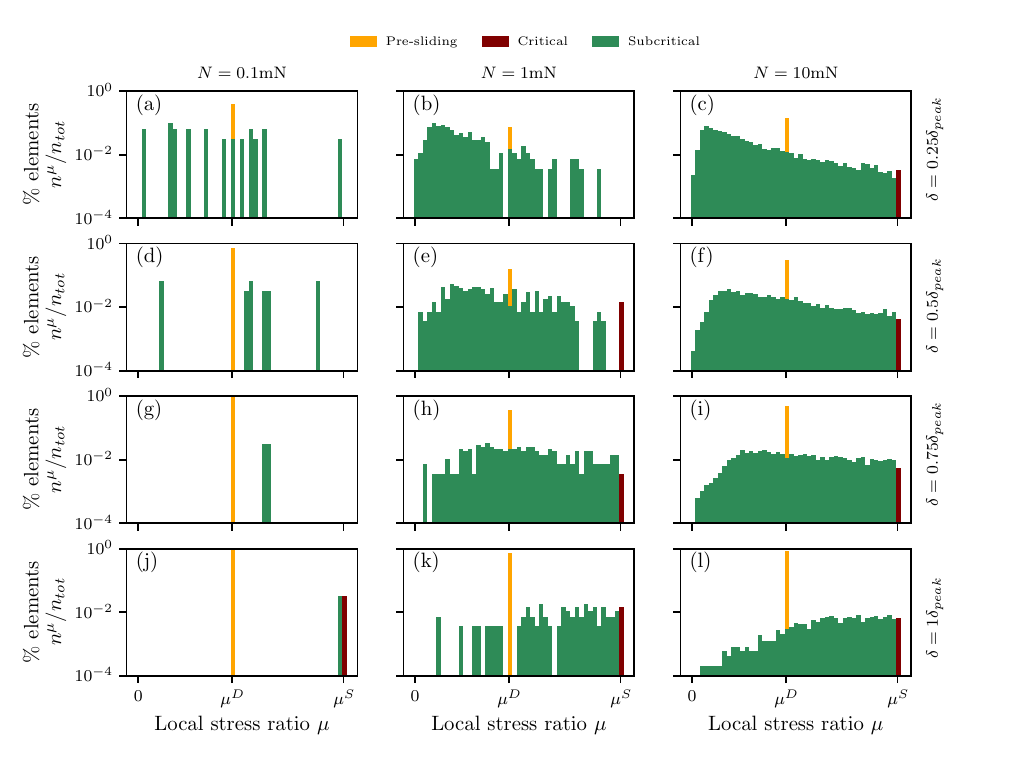}
     \caption{Distribution of the shear-to-normal stress ratio for loads of (a-d-g-j) N=0.1$mN$, (b-e-h-k) $N=1$mN and (c-f-i-l) $N=10$mN. The imposed displacement corresponds to $\delta/\delta_{peak} = 0.25$, $0.5$, $0.75$ and $1$, increasing from the top row to the bottom one.}
    \label{fig:figs6}
\end{figure*}

Each movie shows the evolution of the distribution of local stress ratios with increasing imposed displacement, with the critical asperities ($\mu = \mu^S$) in brown, the pre-sliding asperities ($\mu = \mu^D$) in yellow, and the subcritical asperities ($\mu < \mu^0$) in green. The movies are available \href{https://youtube.com/playlist?list=PLT7c4IN61XQNcKJa2qVdejrUf4y4jP9aG&si=fOAnZ-DrTdHhzNXx}{\color{blue}online}. Snapshots of the movies are shown in Fig.\ref{fig:figs6}.

\clearpage
\newpage